\documentclass[11pt,twoside]{article}
\usepackage{asp2010}

\resetcounters

\begin{document}

\title{Understanding the Angular Momentum Loss of Low-Mass Stars: The Case of V374~Peg}
\author{A.~A.~Vidotto$^1$, M.~Jardine$^1$, M.~Opher$^2$, {J.~F.~Donati}$^{3}$, and {T.~I.~Gombosi}$^{4}$
\affil{$^1$SUPA, U.~St Andrews, North Haugh, St Andrews, KY16 9SS, UK}
\affil{$^2$George Mason University, 4400 University Drive, Fairfax, VA, 22030, USA}
\affil{$^3$LATT - CNRS/U.~de Toulouse, 14 Av.~E.~Belin, Toulouse, F-31400, France}
\affil{$^4$U.~Michigan, 1517 Space Research Building, Ann Arbor, MI, 48109, USA}}

\begin{abstract}
Recently, surface magnetic field maps had been acquired for a small sample of active M dwarfs, showing that fully convective stars (spectral types $\sim$ M4 and later) host intense ($\sim$ kG), mainly axi-symmetrical poloidal fields. In particular, the rapidly rotating M dwarf V374~Peg (M4), believed to lie near the theoretical full convection threshold, presents a stable magnetic topology on a time-scale of $\sim 1$~yr. The rapid rotation of V374~Peg ($P=0.44$~days) along with its intense magnetic field point toward a magneto-centrifugally acceleration of a coronal wind. In this work, we aim at investigating the structure of the coronal magnetic field in the M dwarf V374~Peg by means of three-dimensional magnetohydrodynamical (MHD) numerical simulations of the coronal wind. For the first time, an observationally derived surface magnetic field map is implemented in MHD models of stellar winds for a low-mass star. We self-consistently take into consideration the interaction of the outflowing wind with the magnetic field and vice versa. Hence, from the interplay between magnetic forces and wind forces, we are able to determine the configuration of the magnetic field and the structure of the coronal winds. Our results enable us to evaluate the angular momentum loss of the rapidly rotating M dwarf V374~Peg.
\end{abstract}

\section{Introduction}
The rotational evolution of M dwarf (dM) stars can be inferred from observations of open clusters at different ages \citep{2006MNRAS.370..954I, 2007MNRAS.381.1638S, 2009ApJ...691..342H, 2009MNRAS.400..451, 2009ApJ...695..679M}. In young ($\lesssim 700~$Myr) open clusters, dM stars still present high rotation rates, which suggests that angular momentum losses at the early main-sequence phase are negligible for them \citep{2009IAUS..258..363I}. However, as the cluster ages ($\gtrsim 700~$Myr), the number of rapidly rotating dM stars decreases, implying that there should exist a mechanism of angular momentum removal that acts on time-scales of a few hundred million years \citep{2007MNRAS.381.1638S}. For solar-like main sequence stars, the magnetised stellar wind is believed to spin down the star by carrying away stellar angular momentum. It has been observationally established that the angular velocity rate $\Omega_0$ for solar-like stars varies as a function of age $t$ as $\Omega_0 \propto t^{-1/2}$ \citep{1972ApJ...171..565S}. However, it seems that the empirical \citeauthor{1972ApJ...171..565S}'s law is not valid for low-mass stars, suggesting that a solar-type wind (i.e., with low velocities and mass-loss rates) cannot reproduce the rotational evolution of fully-convective stars. 

The existence of hot coronae, rapid rotation, and high levels of magnetic activity in dM stars suggests the presence of winds with an enhanced mass loss as compared to the solar wind. However, the low-density, optically thin winds of these stars prevents the observation of traditional mass-loss signatures, such as P~Cygni profiles. The still unobserved high mass-loss rates from dM stars could be able to disperse debris discs, explaining why discs around dM stars older than $\gtrsim 10$~Myr are scarcely found \citep{2005ApJ...631.1161P}. Estimates of mass-loss rates from dM stars vary considerably. It has been suggested that the coronal winds of dM stars, despite of being very tenuous, possess mass-loss rates ($\dot{M}$) that can considerably exceed the solar value ($\dot{M}_\odot \simeq 2 \times 10^{-14}~{\rm M}_\odot ~{\rm yr}^{-1}$) by factors of $10$ to $10^4$ \citep{1992ApJ...397..225M, 1992SvA....36...70B, 1996ApJ...462L..91L, 1997A&A...319..578V, 2001ApJ...546L..57W}, although \citet{2001ApJ...547L..49W} claim an upper limit of $\dot{M} \lesssim 4 \times 10^{-15}~{\rm M}_\odot ~{\rm yr}^{-1}$ for Proxima Centauri (dM5.5e), $5$ times below the value of the solar wind mass-loss rate. 

In this work, we investigate the coronal wind of a specific fully-convective dM star, V374~Peg, for which observed surface magnetic maps have been acquired \citep{2006Sci...311..633D, 2008MNRAS.384...77M}. For this, we use three-dimensional (3D) magnetohydrodynamics (MHD) simulations based on our previous models developed for solar-like stars \citep{2009ApJ...699..441V} and weak-lined T Tauri stars \citep{2009ApJ...703.1734V,vidotto10}. For the first time, an observationally derived surface magnetic field map is implemented in MHD models of stellar winds for a low-mass star. V374~Peg is a suitable case for modelling as a first step, because its surface magnetic distribution is close to potential, which implies that the adopted boundary conditions match the observed map closely. We self-consistently take into consideration the interaction of the outflowing wind with the magnetic field and vice-versa. Hence, from the interplay between magnetic forces and wind forces, we are able to determine the configuration of the magnetic field and the structure of its coronal wind. More details of this work can be found in \citet{vidottomn}.

\section{The Numerical Model and Results}

\begin{figure}[!ht]
\plottwo{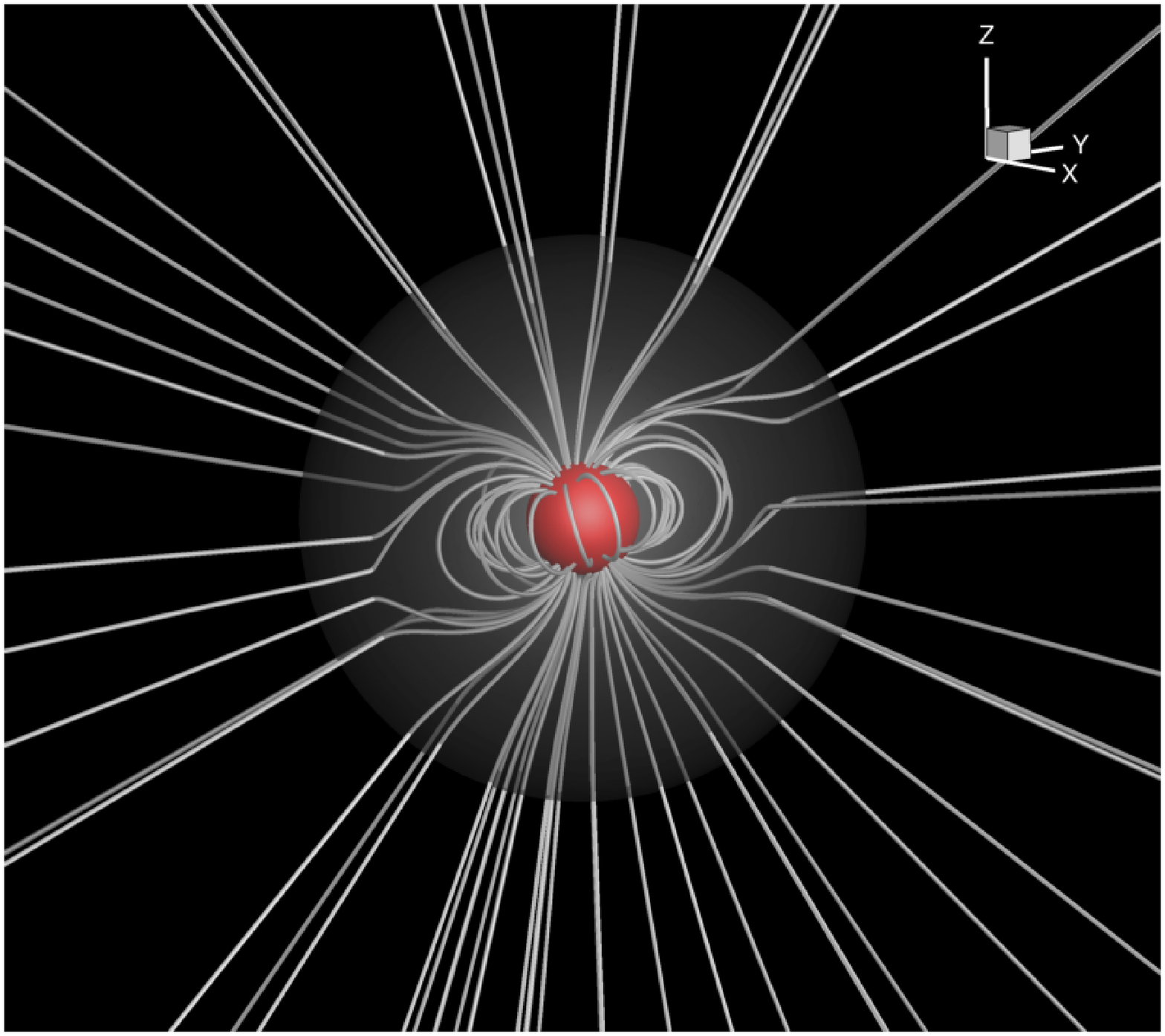}{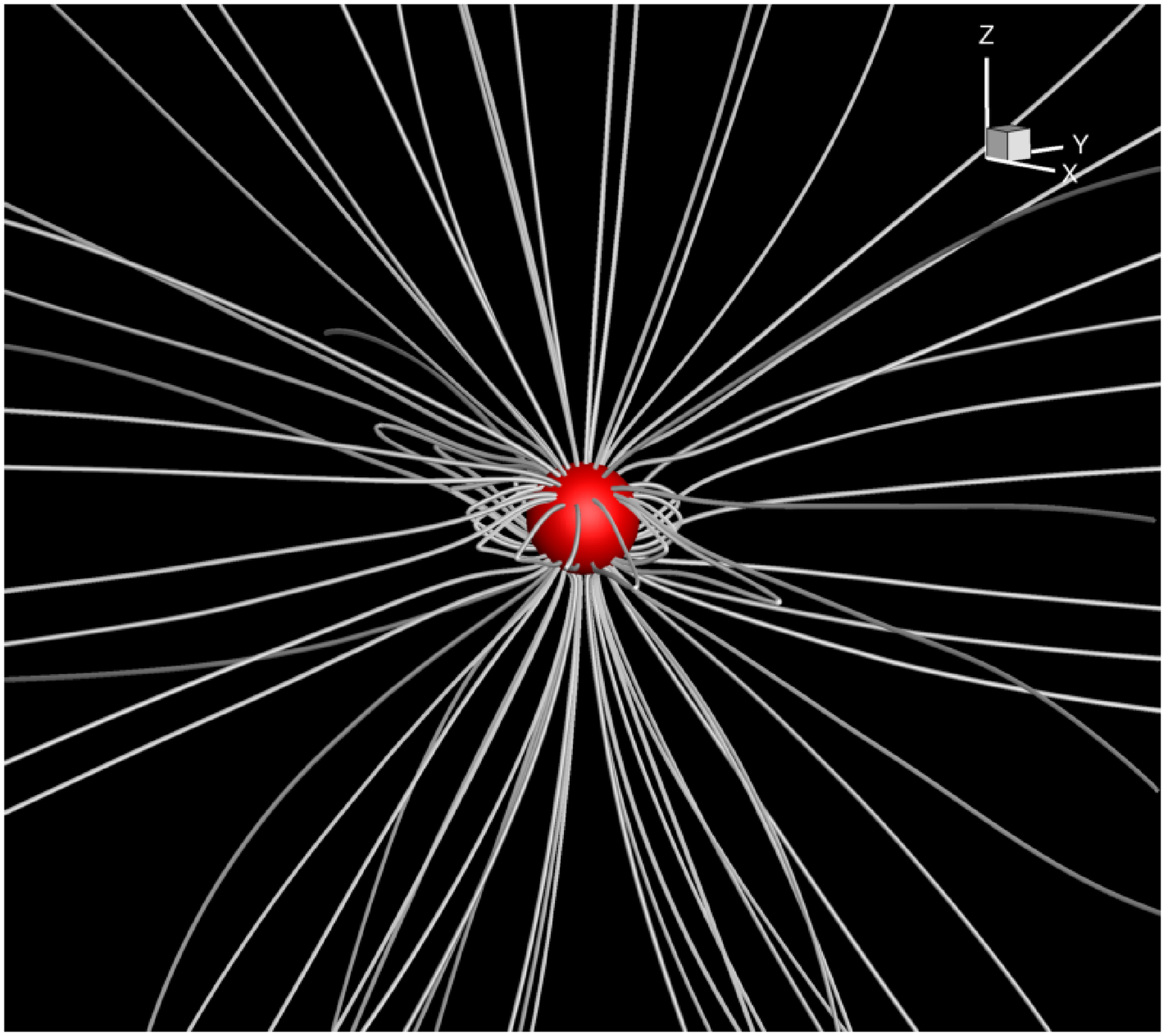}
\caption{(a) Initial configuration of magnetic field lines for the extrapolation of the surface map using the potential-field source surface (PFSS) technique, where the grey circumference represents the position of the source surface. (b) Final configuration of the magnetic field lines after the self-consistent interaction with the stellar wind. \label{fig.IC}}
\end{figure}

 V374~Peg has mass $M_* = 0.28~M_\odot$, radius $R_*=0.34~R_\odot$ and is rotating with negligible differential rotation (i.e., as a solid body) with a period of rotation $P_0=0.44$~d \citep{2008MNRAS.384...77M}. We consider that its axis of rotation lies in the $z$-direction. To perform the simulations, we use the 3D MHD numerical code BATS-R-US developed at University of Michigan \citep{1999JCoPh.154..284P}, which solves the ideal MHD equations. The simulations are initialised with a 1D hydrodynamical wind for a fully ionised plasma of hydrogen. Immersed in this wind we consider a magnetic field anchored on the stellar surface that has a geometry derived from extrapolations from observed surface magnetic maps using the potential-field source surface (PFSS) method (Figure~\ref{fig.IC}a). The MHD solution is evolved in time from the initial magnetic field configuration to a fully self-consistent solution (Figure~\ref{fig.IC}b). We do not adopt fixed topologies for either the magnetic field or for the wind, as both the wind and magnetic field lines are allowed to interact with each other. 

In the PFSS model, the stellar wind plasma is not included directly, but its effects on the magnetic field (and vice-versa) are incorporated through the inclusion of the source surface. Such a surface, for instance, alters the number of open magnetic field lines, through where a stellar wind could escape. The source surface (grey sphere in Figure~\ref{fig.IC}a) is chosen to lie at $r_{\rm SS} = 5~R_*$, where beyond that, the magnetic field is considered to be purely radial. PFSS methods are usually criticised because their basic assumptions (the magnetic field is a potential field and the source surface is spherical) may not always be met. However, the advantage of the PFSS method over the MHD models relies on its simplicity: it is simpler to implement and requires much less computer resources than MHD models. In our model, we use the magnetic field configuration derived by the PFSS method as initial condition and boundary condition at the surface of the star. We note that the surface of the star that occupies co-latitudes $\gtrsim 120^{\rm o}$ is never in view as the star rotates and so the magnetic field there can not be reconstructed reliably.

Our simulations require a set of input parameters for the wind. Unfortunately, some of them are poorly constrained by observations. For V374~Peg, the magnetic field is the better-constrained parameter. We have, therefore, implemented in our previous models \citep{vidotto10} surface magnetic maps derived by data acquired in 2005 Aug \citep{2006Sci...311..633D}. These observations show that V374~Peg hosts an intense, mainly axi-symmetrical dipolar magnetic field, with maximum intensity of about $1660$~G, i.e., $3$ orders of magnitude larger than the Sun. 

The wind temperature and density are less constrained for V374~Peg. We, therefore, adopt values representative of dM stars. dM stars are believed to host coronae with a high-temperature plasma $\sim 10^7$~K in conjunction with a low-temperature one $2$ -- $3 \times 10^6$~K \citep{1990ApJ...365..704S, 1996ApJ...463..707G}. In our models, we adopt a temperature at the base of the coronal wind of $T_0 = 2\times 10^6$~K or $10^7$~K. These coronal temperatures are about the same order of magnitude as the solar coronal temperatures of $1.56 \times 10^6~$K.

Coronal densities inferred from X-ray observations of dM stars suggest densities ranging from $10^{10}$~cm$^{-3}$ to $5 \times 10^{12}$~cm$^{-3}$ \citep{2002A&A...394..911N, 2004A&A...427..667N}. Therefore, we adopt, at the base of the coronal wind, densities in the range $10^{10}$ -- $10^{12}$~cm$^{-3}$. Compared to the solar coronal density of about $\sim 2 \times 10^8$~cm$^{-3}$, coronal densities inferred for dM stars are about $2$ -- $4$ orders of magnitude larger than for the solar corona.

The density, along with the magnetic field, are key parameters in defining the magnetic field configuration of the stellar wind and its velocity profile \citep{2009ApJ...699..441V,2009ApJ...703.1734V}. Together, they define the plasma-$\beta$, defined by the ratio of thermal to magnetic energy densities. Therefore, at the base of the coronal wind of V374~Peg, 
\begin{equation}\label{eq.beta0}
\beta_0 = \frac{n_0 k_B T_0}{B_0^2/(8\pi)} \simeq 2.5 \times 10^{-5} n_{10} T_6, 
\end{equation}
where the index ``0'' means the variable is evaluated at the base of the coronal wind, $n_{10} = n_0/(10^{10}$~cm$^{-3})$ and $T_6=T_0/(2\times10^6$~K). For $n_{10}=1$ and $T_6=1$, $\beta_0$ is about $5$ orders of magnitude smaller than for the solar wind \citep{1971SoPh...18..258P}. This implies that the winds of dM stars are highly magnetised and, therefore, are expected to differ from solar-type winds. 

Table~\ref{table} presents the parameters adopted for the set of simulations we performed.

\begin{table} 
\caption{Adopted parameters for the simulations. The columns are, respectively: the case name, the density $n_0$ and temperature $T_0$ at the base of the coronal wind ($r=R_*$), the plasma-$\beta$ parameter evaluated at $R_*$ [Eq.~\ref{eq.beta0}], the mass loss rate $\dot{M}$, the angular momentum loss rate $\dot{J}$, and the time-scale for rotational braking $\tau$.\label{table}}   
\begin{center}
{\small
\begin{tabular}{ccccccc}  
\tableline   
\noalign{\smallskip}
{Case} &  {$n_0$}  & {$T_0$} &    {$\beta_0$} &  {$\dot{M}$}  & {$\dot{J}$} &   {$\tau$} \\
  & [cm$^{-3}$] & [MK]      &  & [$10^{-11}$~M$_\odot~{\rm yr}^{-1}$] & [$10^{33}$~erg~s$^{-1}$] &  [Myr] \\
\noalign{\smallskip}
\tableline
\noalign{\smallskip}
1Map  &    $10^{10}$ & $2$  &    $2.52\times10^{-5}$ &    $4.2$ & $3.4$ &   $180$    \\
2Map  &    $10^{11}$ & $2$  &    $2.52\times10^{-4}$ &    $14$ & $7.6$ &   $84$   \\
3Map  &    $10^{12}$ & $2$  &    $2.52\times10^{-3}$ &    $50$ & $32$ &   $17$   \\
4Map  &    $10^{11}$ & $10$ &    $1.26\times10^{-3}$ &    $26$ & $9.1$ &   $48$    \\
\noalign{\smallskip}
\tableline          
\end{tabular}
}
\end{center}
\end{table}

We were able to find a MHD solution for the wind for all the simulations we ran, showing that it is possible to develop coronal wind models with a realistic distribution of magnetic field. In general, MHD wind models are studied under the assumption of simplistic magnetic field configurations, especially when in pursuit of an analytical solution. Therefore, the study of a magnetised coronal wind where an observed magnetic field distribution is considered has long been awaited. Our work also sheds some light on the yet unobserved winds from dM stars.

\begin{figure*}
\plottwo{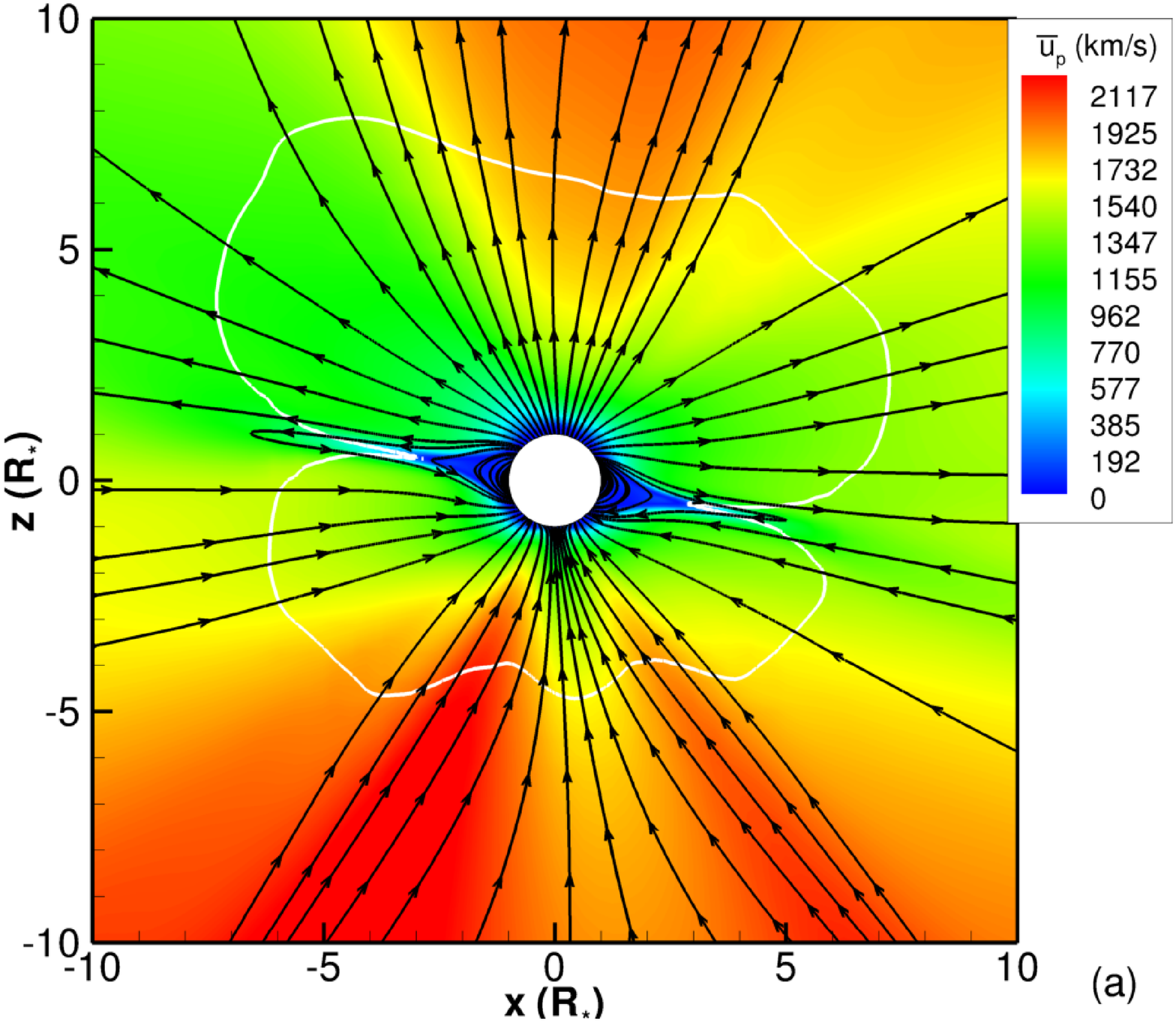}{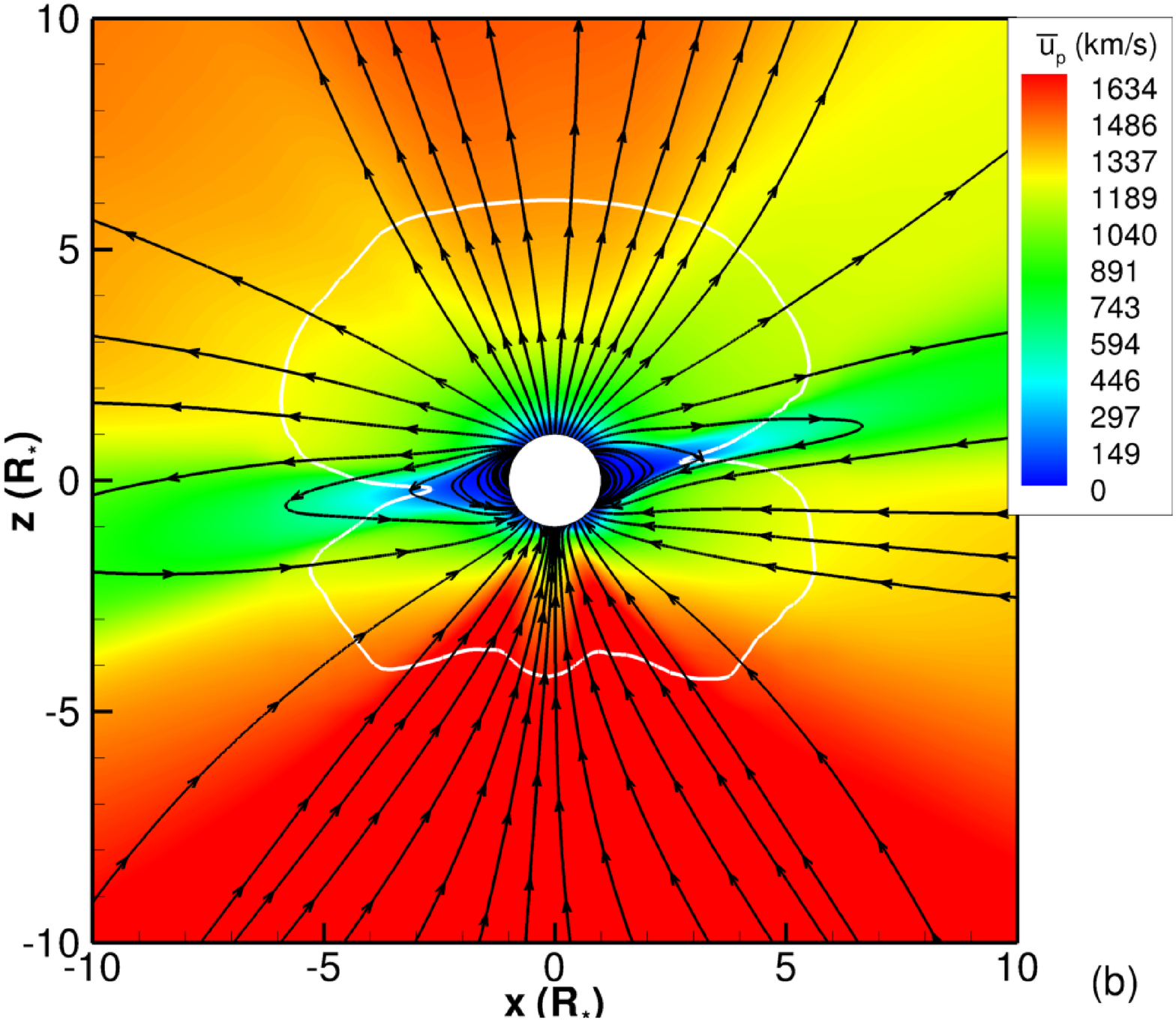}
 \caption{Meridional cut of scaled poloidal wind velocity $\bar{u}_p=u_p/\sqrt{n_{12}}$ profile (a) for cases where a magnetic surface map was used 1Map, 2Map, and 3Map, and (b) 4Map. Black lines represent the magnetic field configuration, and white line is the location of the Alfv\'en surface. \label{fig.windvel}}
\end{figure*}

By comparing cases where only the base coronal density $n_0$ was varied, we found that the poloidal velocity of the wind scales approximately as
\begin{equation}\label{eq.windlaw}
u_p^2 \propto \frac{1}{n_0} {\rm ~for~a~given~}B_0.
\end{equation}
This qualitatively agrees with previous results \citep{2009ApJ...699..441V}, where it was found that an increase in the density leads to winds with lower velocities. Simulations presented here are in a very low-$\beta$ regime, where magnetic effects completely override thermal and kinematic effects of the wind. Therefore, Eq.~(\ref{eq.windlaw}) should be treated with caution, as under different $\beta$ regimes (for example, when it approaches $\beta_0\sim 1$), it becomes invalid. Figure~\ref{fig.windvel}a presents the scaled wind velocity profile $\bar{u}_p$ for cases 1Map, 2Map, and 3Map,
\begin{equation}\label{eq.scaledv}
\bar{u}_p = {u_p}{n_{12}^{1/2}} ,
\end{equation}
where $n_{12} = n_0 /(10^{12}$~cm$^{-3}$). We note that the wind terminal velocity is $u_\infty \approx [1300$ -- $2100] n_{12}^{-1/2}~{\rm km~s}^{-1}$, where the range of velocities refers to different wind latitudes (low-wind velocity near the equator, high-wind velocity around the poles). Because the magnetic field in the lower hemisphere of the star is not reliably reconstructed (co-latitudes $\gtrsim 120^{\rm o}$ of the surface of the star are not observed), a high-velocity wind develops there. Although this feature is local and does not affect the remaining parts of the grid other than radially away from the stellar surface, it is an artifact of our method and should not be taken into consideration (e.g., see the accumulation of magnetic field lines near the low-hemisphere of the star in Figures~\ref{fig.windvel}a and \ref{fig.windvel}b). 

Equation~(\ref{eq.windlaw}) also implies that the mass-loss rate of the wind ($\dot{M} \propto \rho u_r $) should scale as
\begin{equation}\label{eq.windmassloss}
\dot{M} \propto \rho u_r \propto  n_0^{1/2},
\end{equation}
which means that, despite the fact that the wind velocity of case 3Map is $10$ times smaller than case 1Map [Eq.~(\ref{eq.windlaw})], its mass-loss rate is one order of magnitude larger than for case 1Map [Eq.~(\ref{eq.windmassloss})]. This has implications on the efficiency of angular momentum loss, as will be shown later. The mass-loss rates for cases 1Map, 2Map, and 3Map are $\dot{M} \approx 4\times 10^{-10} n_{12}^{1/2}~{\rm M}_\odot ~{\rm yr}^{-1}$. 

Figure~\ref{fig.windvel}b shows the scaled poloidal velocity profile $\bar{u}_p$ for case 4Map. This case considers a different temperature at the base of the corona ($10^7$~K as opposed to $2\times10^6$~K), and, because of that, has a larger $\beta_0$ (Table~\ref{table}). For this case, we did not find an analytical expression relating velocity and temperature. The Alfv\'en surface location and configuration of magnetic field lines are similar to the other dipolar cases, but the scaled wind velocity $\bar{u}_p$ is smaller than the previous cases: $u_\infty \approx [850$ -- $1600] n_{12}^{-1/2}~{\rm km~s}^{-1}$. The lower velocity observed in case 4Map happens because of its higher $\beta_0$. The mass-loss rate for case 4Map is $\dot{M} \simeq 2.6\times 10^{-10}~{\rm M}_\odot ~{\rm yr}^{-1}$.

Overall, our solutions differ considerably from the solar wind solution, where a low-velocity wind (terminal velocities of $u_{\infty , \odot} \simeq 400$ -- $800$~km~s$^{-1}$) with low mass-loss rate ($\dot{M}_\odot \simeq 2 \times 10^{-14}~{\rm M}_\odot ~{\rm yr}^{-1}$) is found. We note that, based on more simplistic wind models, such as \citet{1967ApJ...148..217W}, in the fast magnetic rotator limit, wind terminal velocities of $\simeq 3320~{\rm km~s}^{-1}$ are expected for a wind mass-loss rate of about $10^{-11}~{\rm M}_\odot ~{\rm yr}^{-1}$. 

%

\section{Discussion: Angular Momentum Evolution}
Observations of the rotation evolution of dM stars in open clusters at different ages provide a way to constrain the time-scale $\tau$ for the angular-momentum loss. It has been suggested that $\tau \sim 200$~Myr or, mostly likely $400$ -- $800$~Myr, \citep{2007MNRAS.381.1638S} for dM stars. Angular momentum of the star is carried away by the stellar wind. Because in our simulations there is no axi-symmetry, the torque $\dot{\bf J}$ on the star has $x$, $y$ and $z$ components. Here, we are interested only on the $z$-component, as it is the only one responsible for the rotational braking (because the angular velocity of the star points in the $z$-direction). The $z$-component of the angular momentum carried by the wind is \citep{1970MNRAS.149..197M}
\begin{eqnarray}\label{eq.jdot}
\dot{J} &=&  \left[\alpha{\bf \hat{z}} \times \int_{V_A}  {\bf r} \times \rho ({\bf V }+ \alpha{\bf \hat{z}} \times {\bf r}){\rm d}V_A \right]_z + \int_{S_A} \left( p + \frac{B^2}{8\pi} \right) ({\bf r} \times {\bf \hat{n}})_z {\rm d}S_A \nonumber \\ &+& \int_{S_A} \left[ {\bf r} \times (\alpha{\bf \hat{z}} \times {\bf r})\right]_z \rho {\bf V} \cdot {\bf \hat{n}} {\rm d} { S_A} ,
\end{eqnarray} 
where ${\bf V} = {\bf u} - \alpha {\bf \hat{z}}\times {\bf r}$ is the velocity vector in the frame rotating with angular velocity $\alpha {\bf \hat{z}}$, ${\bf \hat{z}}$ is the unit vector that points in the $z$-direction, $S_A$ is the Alfv\'en surface that delimits the volume $V_A$, and ${\bf \hat{n}}$ is the normal unit vector to the Alfv\'en surface. The first term on the right of Eq.~(\ref{eq.jdot}) does not contribute to the $z$-component torque and is therefore null. The second term disappears in the case of a spherical Alfv\'en surface, but it is non-null in the cases where a surface magnetic map is considered and it becomes relatively more important for the cases with larger adopted $\beta_0$. The third term is the dominant term in our simulations. 

We can estimate the time-scale for rotational braking as $\tau = {J}/{\dot{J}}$, where $J$ is the angular momentum of the star. If we assume a spherical star with a uniform density, then $J = 2/5 M_* R_*^2 \Omega_0$ and the time-scale is
\begin{equation}
\tau \simeq \frac{9 \times 10^{36}}{\dot{J}} \left( \frac{M_*}{M_\odot}\right) \left( \frac{1~{\rm d}}{P_0} \right) \left(  \frac{R_*}{R_\odot}\right)^2 ~{\rm Myr},
\end{equation} 
where $P_0=2\pi/\Omega_0$ is the rotational period of the star. For V374~Peg, this results in 
\begin{equation}\label{eq.tau}
\tau \simeq \frac{6.45 \times 10^{35}}{\dot{J}} ~{\rm Myr}.
\end{equation} 

Because $\dot{J}$ depends on the mass flux crossing a given surface, i.e., on the mass-loss rate of the wind $\dot{M}$, from Eq.~(\ref{eq.windmassloss}), we have a rough scaling relation between $\dot{J}$ and $\dot{M}$ for cases 1Map, 2Map, and 3Map
\begin{equation}\label{windangloss}
\dot{J} \propto \dot{M} \propto n_0^{1/2}  ,
\end{equation} 
which implies in a time-scale [Eq.~(\ref{eq.tau})] for rotational braking that scales as 
\begin{equation}\label{eq.windtimescale}
\tau \propto n_0^{-1/2} .
\end{equation} 
For cases 1Map, 2Map, and 3Map, $\tau \simeq 18 {n_{12}^{-1/2}}$~Myr, well below the estimated solar spin-down time $\tau_\odot \simeq 7$~Gyr \citep{1967ApJ...148..217W}.

Table~\ref{table} presents the mass and angular momentum loss rates, and the time-scale for rotational braking calculated for all simulations, where we verify the approximate scaling given by Eqs.~(\ref{eq.windmassloss}), (\ref{windangloss}), and (\ref{eq.windtimescale}). Comparing to the observationally derived rotational braking time-scales of a couple of hundreds of Myr for dM stars is open clusters \citep{2007MNRAS.381.1638S}, we tend to rule out cases with larger coronal base densities (i.e., $n_0 \gtrsim 10^{11}~{\rm cm}^{-3}$). According to this comparison, the most plausible wind density is the one used for models 1Map. Such a density is also able to reproduce typical emission measures of dM stars (${\rm EM} \approx 10^{51}~{\rm cm}^{-3}$) and comparatively (with the remaining cases) smaller mass-loss rates and higher wind velocities. Ultimately, when the star ages, the stellar rotation brakes, reducing the stellar surface magnetic field intensity, and therefore the wind velocity. 

With the inclusion of an observed distribution of surface magnetic field, we head towards a more realistic modelling of magnetised coronal winds. Never the less, our model presents limitations, such as the neglect of a detailed energy balance. Instead, we consider a polytropic relation between pressure and density parametrised through $\gamma$ in the derivation of the energy equation of the wind. Once the magnetic field distribution is set, the thermal pressure adjusts itself in order to provide a distribution of heating/cooling that is able to support the MHD solution obtained \citep{1986ApJ...302..163L}. If the wind of V374~Peg is able to cool down, e.g., by radiative cooling not considered in our models, the terminal velocities of the wind could be considerably smaller. Depending on where in the wind energy deposition (or removal) occurs, the wind velocity may change, without affecting the mass-loss rates. For instance, if a substantial cooling occurs above the Alfv\'en surface, the velocity profile of the wind from that point outwards will be affected. As the information of what is happening above the Alfv\'en point cannot be transmitted to the sub-Alfv\'enic region, the wind density and velocity profiles in the proximity of the star will not be changed, and consequently neither the stellar mass-loss/angular momentum-loss rates.  

\acknowledgements The simulations presented here were performed at Columbia (NASA Ames).

\end{document}